\documentclass[aps,pre,twocolumn,showpacs]{revtex4}
\usepackage[dvips]{graphicx}
\usepackage{amsmath}
\usepackage{times}

\def \rmin{r_{\mathrm{min}}}
\def \rmax{r_{\mathrm{max}}}

\begin{document}
\author{ %
Dmitry V. Savin and Hans-J\"urgen Sommers}
\affiliation{ %
Fachbereich Physik, Universit\"at Duisburg-Essen, 45117 Essen, Germany}

\title{Distribution of reflection eigenvalues in many-channel
chaotic cavities with absorption}
\date{November 12, 2003 }%

\begin{abstract}
The reflection matrix $R=S^{\dagger}S$, with $S$ being the scattering matrix,
differs from the unit one, when absorption is finite. Using the random matrix
approach, we calculate analytically the distribution function of its
eigenvalues in the limit of a large number of propagating modes in the leads
attached to a chaotic cavity. The obtained result is independent on the
presence of time-reversal symmetry in the system, being valid at finite
absorption and arbitrary openness of the system. The particular cases of
perfectly and weakly open cavities are considered in detail. An application
of our results to the problem of thermal emission from random media is briefly
discussed.
\end{abstract}


\maketitle

\textit{Introduction.--} %
When absorption in a cavity is finite, part of the incoming flux gets
irreversibly lost in the walls, breaking thus the unitarity of the scattering
matrix. The mismatch between incoming and outgoing fluxes in the scattering
process can be naturally described by means of the reflection matrix
$R=S_{\gamma}^{\dagger}S_{\gamma}$, where $S_{\gamma}$ denotes the subunitary
(at nonzero absorption) scattering matrix to be precisely defined below. The
matrix $R$ has the positive eigenvalues $r_c\leq1$, the so-called
\textit{reflection eigenvalues}. Their statistical properties in chaotic
resonance scattering attract much attention presently, both experimentally
\cite{Doron1990,Mendez-Sanchez2003} and theoretically
\cite{Bruce1996,Beenakker1996,Beenakker1998,Kogan2000,Ramakrishna2000,%
Savin2003i,Fyodorov2003i}, using the random-matrix theory
approach \cite{Verbaarschot1985,Beenakker1997}.

In the limit of weak absorption, $R$ was found
\cite{Doron1990,Ramakrishna2000} to be related to the Smith's time-delay
matrix at zero absorption those statistical properties were extensively
studied in recent time
\cite{Lehmann1995b,Fyodorov1997,Savin2001,Brouwer1997,Sommers2001}. Such an
analysis was recently generalized by us to the case of arbitrary absorption
\cite{Savin2003i}, see also \cite{Fyodorov2003i} for the related study. In
the particular case of the single-channel cavity, obtained results explain
partly the recent experiment \cite{Mendez-Sanchez2003} on the reflection
coefficient distribution.

The opposite case of a large number of equivalent channels plays an important
role in chaotic scattering, corresponding to the semiclassical limit of
matrix models \cite{Lewenkopf1991,Lehmann1995a}. In this case  $N$ strongly
overlapping resonances (poles of the $S$-matrix) are excited in the open
cavity through the scattering channels, whose number $M$ scales with $N$. A
new energy scale appears then in addition to the mean level spacing $\Delta$
of the closed cavity \cite{Lehmann1995a,Lehmann1995b}: the empty gap between
the real axis in the complex energy plane and the cloud of the resonances in
its lower half plane. In the physically justified limit $M/N{\ll}1$
\cite{Lehmann1995b} (though both $M,N{\to}\infty$) considered below this gap
is given by the well-known Weisskopf width $\Gamma_W=MT\Delta/2\pi$, with
$T\leq1$ being the transmission coefficient \cite{Lewenkopf1991}. In the
presence of finite absorption the widths of the resonances acquire an
additional contribution, the absorption width $\Gamma_a$ \cite{Savin2003i}.
The two competing processes, the escape into continuum and dissolution into
the walls, are therefore characterized by the ratio
$\Gamma_a/\Gamma_W=\gamma/T$, where the convenient dimensionless parameter
$\gamma\equiv2\pi\Gamma_a/\Delta{M}$ measures the absorption strength.

In this paper we derive the probability distribution function
$\mathcal{P}(r)=M^{-1}\sum_{c=1}^M\overline{\delta(r-r_c)}$ of reflection
eigenvalues, with the bar indicating the statistical average, in the limit
of the large number of channels. This function has, in particular, an
important application for the statistics of thermal emission from random
media and is known only at perfect coupling ($T{=}1$) \cite{Beenakker1998}. 
Here, we calculate it at arbitrary $T$.

Our starting point is the general relation established in
Ref.~\cite{Savin2003i} between $R$ and the effective non-Hermitian
Hamiltonian $\mathcal{H}=H-(i/2)VV^{\dagger}$ of the open system, with the
Hermitian part $H$ standing for the closed counterpart and the amplitudes
$V_n^c$ describing the coupling between $N$ intrinsic and $M$ channel states.
It is achieved by making use of the equivalence
\cite{Kogan2000,Ramakrishna2000,Savin2003i} of uniform absorption to the
pure imaginary shift $E_{\gamma}\equiv{E}+\frac{i}{2}\Gamma_a$ of the
scattering energy $E$, thus giving
$S_{\gamma}\equiv{S}(E_{\gamma})=1-iV^{\dagger}(E_{\gamma}-\mathcal{H})^{-1}V$.
This leads to the following connection between the reflection matrix $R$ and
the time-delay matrix with absorption $Q_{\gamma}$ ($\hbar=1$)
\cite{Savin2003i}:
\begin{equation}\label{R}
R = 1 - \Gamma_a Q_{\gamma} \,,
\end{equation}
where $Q_{\gamma}\equiv Q(E_{\gamma})=
V^{\dagger}[(E_{\gamma}-\mathcal{H})^{\dagger}]^{-1}(E_{\gamma}-\mathcal{H})^{-1}V$.
The distribution function $\mathcal{P}(r)$ is therefore directly related as
\begin{equation}\label{P(r)}
\mathcal{P}(r) = \gamma^{-1}\mathcal{P}_{\tau}\left(\gamma^{-1}(1-r)\right)
\end{equation}
to the distribution
$\mathcal{P}_{\tau}(\tau)=M^{-1}\sum_{c=1}^M\overline{\delta(\tau-Mq_c/t_H)}$
of the proper delay times (eigenvalues $q_c$ of $Q_{\gamma}$) scaled with $M$
and measured in units of the Heisenberg time $t_H=2\pi/\Delta$.

The function (\ref{P(r)}) is normalized to unity. The average reflection
coefficient $\langle{r}\rangle=M^{-1}\overline{\mathrm{tr\,}R}$ is given by
\begin{equation}\label{<r>}
\langle{r}\rangle = \int_0^1dr\,r\,\mathcal{P}(r) =
\frac{T+\gamma(1-T)}{T+\gamma} \,.
\end{equation}
This result follows from the connection \cite{Savin2003i} between
$\langle{r}\rangle$ and the ``norm-leakage'' decay function $P(t)$ introduced
in \cite{Savin1997}: $\langle{r}\rangle=1-\gamma[1-\Gamma_a\int_0^{\infty}dt
e^{-\Gamma_a{t}}P(t)]$. In the limit considered $P(t)$ reduces to the simple
exponent $e^{-\Gamma_W{t}}$, since the widths ceased to fluctuate
\cite{Lehmann1995a,Savin1997}. Below we will also derive Eq.~(\ref{<r>})
using a different method.

\textit{The saddle-point equation.--} %
The scaling limit studied is far from being trivial and can not be obtained
from the known expressions for $\mathcal{P}(r)$ at finite $M$
\cite{Savin2003i} simply by letting $M{\to}\infty$ there. One needs to start
from the original definition that determines the seeking distribution through
the jump of the resolvent
$G(t)=M^{-1}\overline{\mathrm{tr\,}(t-\Gamma_aQ_{\gamma})^{-1}} \equiv
1/t+(\gamma/t^{2})K(t)$ along the discontinuity line $t>0$ as follows:
\begin{equation}\label{P-K}
\mathcal{P}(r)=\frac{\gamma}{\pi(1-r)^2}\mathrm{Im\,}K(t-i0)\big|_{t=1-r} \,.
\end{equation}
Following Refs.~\cite{Savin2003i,Sommers2001}, one can obtain the exact (at
$N{\to}\infty$) representation for $K$, which is convenient for our purposes,
\begin{equation}\label{K}
K(t) =
\left\langle\frac{1}{8}\,\mathrm{str}\left[\left(\Lambda-i\frac{1-t/2}{\sqrt{1-t}}
\Lambda_1\right)k \hat{\sigma}\right]\right\rangle_{\mathcal{L}} \,,
\end{equation}
with the shorthand $\langle(\ldots)\rangle_{\mathcal{L}}=
\int\!d\mu(\hat{\sigma})(\ldots)e^{(M/2)\mathcal{L}}$ and
\begin{eqnarray}\label{L}
\mathcal{L} &=& (\gamma/t)\,\mathrm{str}[(1-{t/2})\Lambda\hat{\sigma}
-i\sqrt{1-t}\Lambda_1\hat{\sigma}] \nonumber\\
&&-\mathrm{str}\ln[1+(2g)^{-1}(\Lambda\hat{\sigma}+\hat{\sigma}\Lambda)]\,,
\end{eqnarray}
as the integral over the noncompact saddle-point manifold of the $8{\times}8$
supermatrix $\hat{\sigma}$ subject to the constraint $\hat{\sigma}^2=1$.
$\Lambda,\Lambda_1$ appearing above are the supermatrix analog of the Pauli
matrices $\sigma_3,\sigma_1$ and $k=+1(-1)$ in the space of commuting
(anticommuting) variables. Definitions of the superalgebra as well as
explicit parametrization, which depends on whether TRS is preserved or
broken, can be found in \cite{Verbaarschot1985,Efetov1996}. At last, the
constant $g=2/T-1\geq1$ is related to the transmission coefficient $T$.

In the limit $M{\to}\infty$ the integration over $\hat{\sigma}$ can also be
done in the saddle-point approximation. Function (\ref{K}) is given by a
saddle-point value of $\hat{\sigma}$, which is found, as usual, by equating
the variance $\delta\mathcal{L}$ to zero. The structure of the latter equation
is determined by that of $\mathcal{L}$. A careful analysis shows that,
independently on TRS, the supermatrix $\hat{\sigma}$ reduces (and so does the
corresponding algebra) in the saddle-point to the $2{\times}2$ \textit{usual}
matrix $\sigma=a\sigma_3+b\sigma_1$  with the imposed constraint $a^2+b^2=1$.
We find that $K=a-ib(1-t/2)/\sqrt{1-t}$, whereas the equation
$\delta\mathcal{L}=0$ gives $(\gamma/t)\sqrt{1-t}K+ib/(g+a)=0$. Eliminating in
this equation $a$ and $b$, one arrives finally after some algebra at the
following saddle-point equation
\begin{equation}\label{1/t}
\frac{1}{t}=\frac{1}{2}-\frac{1}{\gamma K}+
\frac{(K+g)/2}{\pm\sqrt{(K+g)^2+1+4/\gamma^2-g^2}-2/\gamma}\,,
\end{equation}
where the sign ``$+$'' must be taken. Independently on the taken choice of
the sign, equation (\ref{1/t}) can be further equivalently represented as the
following forth order equation in $K$:
\begin{eqnarray}\label{Keq}
\frac{K^4+2gK^3}{t}(1-\frac{1}{t}) &+& \frac{K^2}{4}
\bigl[(g+\frac{2}{\gamma})^2-\frac{8g}{t\gamma}-(1-\frac{2}{t})^2\bigr]
\nonumber\\
&+& \frac{K}{\gamma}(1-\frac{2}{t}) - \frac{1}{\gamma^2} = 0 \,.
\end{eqnarray}
In the zero absorption limit, $\gamma{\to}0$ with fixed $t/\gamma=\tau$, this
equation simplifies to the cubic one found earlier \cite{Sommers2001}.

The choice of the sign ``$+$'' is justified by two reasons. First, only then
the complex solution of Eq.~(\ref{1/t}) yields the distribution
$\mathcal{P}(r)$ nonzero at positive $r$. Second, the function $K(t)$, as
follows directly from its definition, must have at $t{\to}{-\infty}$ the
positive limit $K_{\infty}=t_H^{-1}\overline{\textrm{tr\,}Q_{\gamma}}<1$
\cite{Savin2003i}, which is the mean scaled proper delay time
$\langle\tau\rangle$. The equation for the latter follows then readily from
(\ref{Keq}) as
$[(g+2/\gamma)^2-1]\langle\tau\rangle^2+4\langle\tau\rangle/\gamma-4/\gamma^2=0$
that gives $\langle\tau\rangle=T/(\gamma+T)$, reproducing thus
Eq.~(\ref{<r>}) by virtue of $\langle{r}\rangle=1-\gamma\langle\tau\rangle$.

The higher moments of the distribution can be easily calculated in the same
way, exploiting the $1/t$--expansion further. For example, substituting
$K=\langle\tau\rangle+\gamma\langle\tau^2\rangle/t+O(t^{-2})$ in (\ref{Keq})
and equating there the $1/t$--term to zero, we find
$\langle\tau^2\rangle=\langle\tau\rangle^2[\gamma^2+2(\gamma+T)]/(\gamma+T)^2$.
That yields the variance of reflection eigenvalues as follows:
\begin{equation}\label{var}
\langle{r^2}\rangle-\langle{r}\rangle^2
= \frac{\gamma^2T^2(2\gamma+2T(1-\gamma)-T^2)}{(\gamma+T)^4} \,.
\end{equation}
Remarkably, the variance and, therefore, fluctuations of $r$ are suppressed
in the both limits of weak and strong absorption.

\begin{figure}[b]
\includegraphics[width=0.48\textwidth]{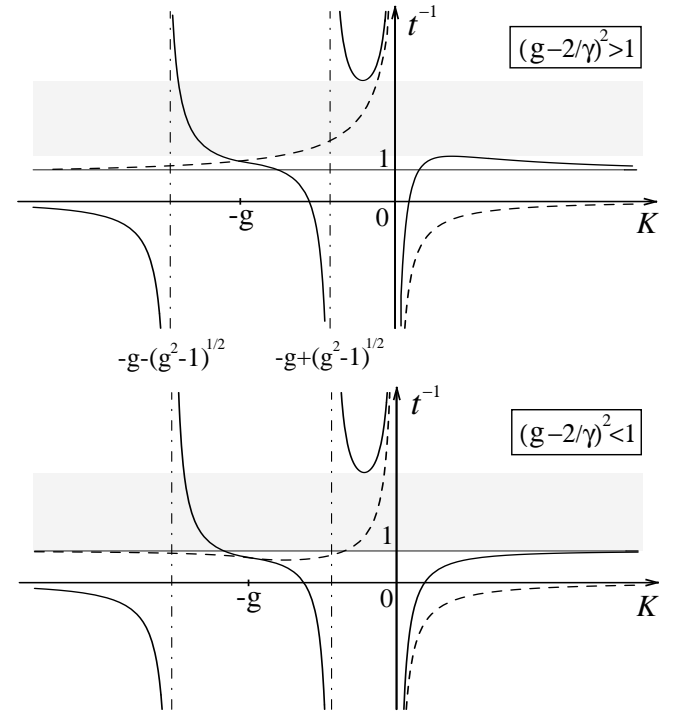}
\caption{The schematic plot of $1/t$ as a function of $K$. The solid and
dashed lines correspond to the choices of ``$+$'' and ``$-$'' sign in
Eq.~(\ref{1/t}), respectively. Altogether they form the complete set of
solutions of the forth-order equation (\ref{Keq}). In the shadowed region the
latter equation has two complex-conjugated roots (and two real roots
irrelevant for our purposes), with $t^{-1}>1$ ($r>0$). Thus, only in this
region the  corresponding density is nonzero. Another solution with
$0<t^{-1}<1$ is unphysical.}
\end{figure}
To understand qualitatively the structure of the solutions of the
saddle-point equation, it is instructive to consider the behavior of $1/t$ as
a function of $K$ shown schematically on Fig.~1. We readily see that the
distribution function $\mathcal{P}(r)$ is nonzero only in the finite domain
$\rmin<r<\rmax$. The value of the upper border $\rmax<1$ at any $g$ and
$\gamma$. In a vicinity of $\rmax$ the distribution behaves as
$\mathcal{P}(r)\propto\sqrt{\rmax-r}$. For the lower border we find $\rmin=0$
and $\mathcal{P}(r)\propto1/\sqrt{r}$, when $(g-2/\gamma)^2\leq1$ or
absorption being from the interval $T\leq\gamma\leq T/(1-T)$, and $\rmin>0$
with $\mathcal{P}(r)\propto\sqrt{r-\rmin}$ otherwise \cite{borders}. Further
analytical study is possible in the following particular cases, which we
consider in detail now.

\textit{Perfect coupling, $g=T=1$.} %
In this case, the region between the two vertical dash-dotted lines on Fig.~1
shrinks to the line $K=-1$. Eliminating the resulting common factor $(K+1)$,
one gets from Eq.~(\ref{Keq}) the following cubic equation
\begin{eqnarray}\label{Keq0}
K^3 + K^2 + \frac{(1-r)^2}{r\gamma^2}
\Bigl[K\bigl(\frac{2\gamma}{1-r}-1-\gamma\bigr)+1\Bigr]=0\,,
\end{eqnarray}
written already in the variable $r=1-t$. We find that the discriminant
$D=[(1+\gamma)/(3\gamma^2r)]^3(r_{+}-r)(r-r_{-})(1-r)^4$ of this equation,
with
\begin{equation}\label{r_pm}
r_{\pm} = \frac{8+20\gamma^2-\gamma^4 \pm
          \gamma(8+\gamma^2)^{3/2}}{8(1+\gamma)^3} \,,
\end{equation}
is positive at $r>0$ (and thereby $\mathrm{Im}K$ is nonzero) only in the
domain $\textrm{max}[0,r_{-}]<r<r_{+}$. This sets explicitly the exact
borders for the searched distribution. $\mathcal{P}(r)$ can be found from 
(\ref{Keq0}) by
 applying Cardan's formulae, reproducing exactly the result of
Ref.~\cite{Beenakker1998} obtained by a different method. 
When absorption is weak, $\gamma<1$, the behavior
of $\mathcal{P}(r)$ can be well approximated by the following simple
interpolation expression \cite{interpol}:
\begin{equation}\label{P0weak}
\mathcal{P}_{\gamma<1}(r) = \frac{C_{\gamma}}{2\pi}
\frac{\sqrt{(r_{+}-r)(r-r_{-})}}{(1-r)^2\sqrt{r}}\,,
\end{equation}
with $C_{\gamma}=1+\frac{3}{2}\gamma+O(\gamma^2)$ being the normalization
constant. In the opposite case of strong absorption, the distribution is
found to be close to
\begin{equation}\label{P0strong}
\mathcal{P}_{\gamma>1}(r) = \frac{2\sqrt{1-r_{+}}}{\pi r_{+}}
\frac{\sqrt{r_{+}/r-1}}{(1-r)^2}\,.
\end{equation}
The limiting distribution (\ref{P0weak}) or (\ref{P0strong}) becomes
asymptotically exact as $\gamma$ diminishes or grows, respectively. All these
features are illustrated on Fig.~2a.
\begin{figure}[b]
\includegraphics[width=0.48\textwidth]{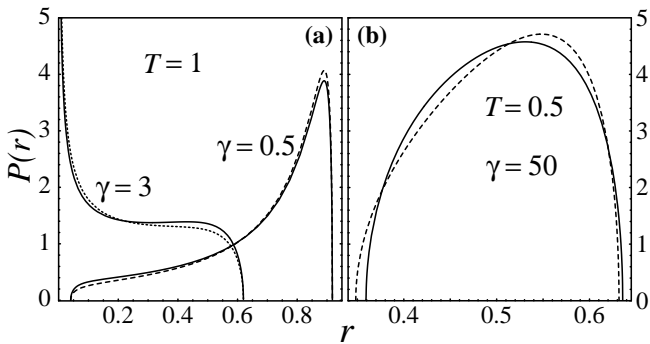}
\caption{The exact reflection distribution (solid lines) compared to the
corresponding approximate expressions (\ref{P0weak}), (\ref{P0strong}), and
(\ref{Pgamma}) at perfect ($T{=}1$, left) and non-perfect ($T{=}0.5$,
right) coupling. }
\end{figure}

\textit{Special coupling, $g^2=1+4/\gamma^2$.} %
Under this condition Eq.~(\ref{1/t}) reduces to a quadratic equation in $K$,
giving readily
\begin{equation}\label{P_g}
\mathcal{P}_{g}(r) = \frac{2\sqrt{1-r_{g}}}{\pi r_{g}}
\frac{\sqrt{r_{g}/r-1}}{(1-r)^2}\,,
\end{equation}
with $r_{g}=2/(1+\sqrt{1+\gamma^2/4})=2/(1+g/\sqrt{g^2-1})$.

\textit{Weak coupling, $g\approx2/T\gg1$.} %
One can find approximate expressions for $\mathcal{P}(r)$, making use of the
scaling $K{\sim}O(1/g)$ in the exact equations. When absorption is weak we
may expand the square root in (\ref{1/t}) and keep only the main contribution
$(2g/\gamma)(1+2gK)^{-1}$ instead of the last term there. That leads again to
a quadratic equation in $K$, giving readily the behavior
$\mathcal{P}(r)\approx(\gamma/2g\pi^2)^{1/2}\sqrt{\rmax-r}/(1-r)^2$ near the
upper border $\rmax\approx1-\gamma/8g$. For $\gamma$ from the small interval
$[T,T/(1-T)]$, when $\rmin{=}0$, $\mathcal{P}(r)$ is given in the leading
approximation by Eq.~(\ref{P_g}), with $r_{g}$ replaced by $\rmax$, because
the applicability condition for (\ref{P_g}) is now satisfied up to
$O(g^{-2})$. At other values of small $\gamma$ there is no reliable
approximation for the lower border available and $\rmin>0$ is to be
found from the general equation \cite{borders}. At last, for very strong
absorption and arbitrary non-perfect coupling, $T\neq1$, we can use scaling
$K{\sim}O(1/\sqrt{\gamma})$ in (\ref{1/t}). That allows us to replace there
the last term with $\frac{1}{2}(K+g)/(1+gK-2/\gamma)$, yielding finally
\begin{equation}\label{Pgamma}
\mathcal{P}_{\gamma\gg1}(r) = \frac{C}{2\pi}
\frac{\sqrt{(r_0+\Delta{r}-r)(r-r_0+\Delta{r})}}{(1-r)^2(1-T+r)}\,,
\end{equation}
with the normalization constant  $C\approx\gamma-2(1-T)$ and
$$r_0 \approx 1-T-\frac{2T(2-3T)}{\gamma}\quad\mathrm{and}\quad
\Delta{r} \approx 2\sqrt{2}T\sqrt{\frac{1-T}{\gamma}}\,,$$
that is valid when $r_{0}\gg\Delta{r}$ or $\gamma\gg8T^2/(1-T)$.
Expression (\ref{Pgamma}) becomes asymptotically exact as $\gamma$ grows,
approaching the ``semi-circle'' distribution with the center at $r_0$ and the
radius $\Delta{r}$, see Fig.~2b.

\textit{Thermal emission.--} %
As an application of our results we consider thermal emission from random
media. In his seminal paper Beenakker \cite{Beenakker1998} has shown that the
quantum optical problem of the photon statistics can be reduced to a
computation of the $S$-matrix of the classical wave equation. In particular,
chaotic radiation may be characterized by the effective number
$\nu_{\textrm{eff}}$ degrees of freedom as follows:
$\nu_{\textrm{eff}}/\nu=(1-\langle{r}\rangle)^2/\langle(1-r)^2\rangle\leq1$
\cite{Beenakker1998}, with $\nu_{\textrm{eff}}=\nu$ for blackbody radiation
\cite{Mandel1995}. We find using (\ref{var}) that at any $\gamma$ and $T$
\begin{equation}\label{nu}
\frac{\nu_{\textrm{eff}}}{\nu}
= \frac{\langle\tau\rangle^2}{\langle\tau^2\rangle}
= \frac{(\gamma+T)^2}{\gamma^2+2(\gamma+T)} \,,
\end{equation}
and a mean photocount $\bar{n}=\nu f\gamma T/(\gamma+T)$, with $f$ being a
Bose-Einstein function. The earlier result \cite{Beenakker1998} is reproduced
at $T{=}1$. Upon the substitution of $\gamma$ with $-\gamma$ one can use
\cite{Beenakker1998} $\bar{n}$ and $\nu_{\textrm{eff}}/\nu$ (\ref{nu}) even
for amplified spontaneous emission below the laser threshold, $\gamma<T$. In
this case our mean photocount agrees with the findings of
Ref.~\cite{Hackenbroich2001}, where the general theory of photocount
statistics in random amplifying media was developed. In the limit of
vanishing absorption or amplification, the ratio $\nu_{\textrm{eff}}/\nu$ is
$T/2$. The large $\gamma$ expansion
$\nu_{\textrm{eff}}/\nu=1-2(1-T)/\gamma+O(\gamma^{-2})$ shows that the
saturation to the blackbody limit gets slower when transmission $T<1$.

%
In conclusion, for many-channel chaotic systems we have derived the general
distribution of reflection eigenvalues at arbitrary values of absorption and
transmission. We note that due to a duality relation
\cite{Beenakker1998,Paasschens1996}, an amplifying system in the linear
regime ($\Gamma_a<\Gamma_W$) is directly linked to the dual absorbing one
through the change of the sign of $\Gamma_a$ in (\ref{R}) and correspondingly
thereafter. As a result, the reflection matrices (and their eigenvalues) of
dual systems are each other's reciprocal. Therefore, the analysis presented
can straightforwardly be extended to the case of linear amplification that
might be also relevant for the rapidly developing field of random lasers
\cite{Beenakker1996,Beenakker1998,Hackenbroich2001,Paasschens1996,Cao2003}.

We are grateful to G. Hackenbroich and C. Viviescas for useful discussions.
The financial support by the SFB/TR 12 der DFG is acknowledged with thanks.


\begin{thebibliography}{23}
\bibitem{borders}
Equation (\ref {Keq}) has two real and one pair of complex
conjugated roots only when the corresponding cubic resolvent has also complex
roots. Equating the discriminant of the latter to zero yields a forth order
equation in $t$, which can be solved, as well as (\ref {Keq}), making use of
computer algebraic packages. The pair of its real solutions determines thus
the borders exactly.
\bibitem{interpol}
The square root in the denominator here $\sqrt{r}\approx\frac{1}{2}(1+r)$
for $r$ close to 1, which could be obtained from a quadratic approximation
of Eq.~(\ref{Keq0}) when $\gamma\ll1$. It is numerically found that $\sqrt{r}$
gives the correct tendency at moderate $\gamma\approx1$, i.e. an enhancement of
$\mathcal{P}(r)$ at smaller $r$.
\end{thebibliography}

\end{document}